
\magnification=\magstep1
\overfullrule=0pt
\def\P{\cal P}

\def\Buildrel#1\over#2{\mathrel{\mathop{\kern0pt #1}\limits_{#2}}}
\quad
\hfill
HD-THEP 95-14
\vskip1.5cm
\centerline{\bf A NOVEL LOOK AT THE MICHAEL LATTICE SUM RULES}
\vskip1cm
\centerline{Heinz J. Rothe}
\medskip
\centerline{Institut f\"ur Theoretische Physik}
\centerline{Universit\"at Heidelberg}
\centerline{Philosophenweg 16, D-69120 Heidelberg}
\vskip2cm
\centerline{\bf Abstract}
\bigskip\bigskip\bigskip
\def\truepageno{\footline={\hss\tenrm\folio\hss}}
\nopagenumbers
\baselineskip20pt
\noindent
We reconsider the derivation of the Michael lattice sum rules, which relate the
energy and action stored in a flux tube of a quark-antiquark pair to the
static interquark
potential, and show that they
require essential corrections.
We then find, using the coupling constant sum rule of Karsch, that the total
Minkowski field
energy does not match the interquark potential, if one follows conventional
notions.
The implications of this result are discussed.
\vfill\eject
\pageno = 1 \truepageno
\baselineskip20pt
Monte Carlo simulations of non-abelian gauge theories are usually carried out
on a lattice with equal lattice spacings in the spatial and euclidean
time directions. A
lattice regularization using different cutoffs for the space and time
directions
must of course yield the same results for physical observables in the continuum
limit. A knowledge of the corresponding regularized action is required
to relate thermodynamical
observables to expectation values of lattice operators, since such
a regularization
allows one to vary independently the temperature and volume of the
system [1-3]. The action
then depends on two coupling
constants associated with the temporal and space-like plaquettes.
These couplings are functions of the spatial lattice cutoff
and the anisotropy parameter $\xi$, defined as the ratio of the
spatial to temporal lattice spacing [2,3]. When taking the continuum limit
with the anisotropy parameter held fixed,
physical observables, such as the interquark potential and particle
masses, should not depend on $\xi$. By requiring
that the string tension obtained either from the expectation value of a
space-time or
a space-like Wilson loop be invariant under changes in the anisotropy
parameter,
Karsch [3] was able to show that the sum of the
derivatives of the inverse coupling constants squared
with respect to the anisotropy parameter, evaluated on an isotropic lattice,
is given
in the continuum limit
by the first coefficient in the perturbative expansion of the $\beta$-function.
Furthermore, by requiring that in the continuum limit the effective
action be independent of the lattice regularization chosen, he was able to
determine
the dependence of the couplings on the anisotropy parameter.
\smallskip
The formulation of $SU(N)$ gauge theories on an anisotropic lattice
has been used in [4] to derive sum rules relating the potential of a
quark-antiquark
pair to correlators of the action, or of the chromoelectric and chromomagnetic
field
energy, with the Wilson loop. These lattice sum rules are known as Michael's
sum rules.
Comparing the action sum rule given in [4] with
the corresponding sum rule in ref. [3], one finds that they disagree by
a factor of 2. That the action sum rule in [4] is actually incorrect has
been noted
recently in [5], where
the authors derive this sum rule within the framework of the continuum
formulation. Motivated by these observations, we have reexamined the
derivation of the Michael
sum rules, and find that there are important corrections to both the action and
the energy sum rule of ref. [4]. Using the coupling constant sum rule
of Karsch [3], we
then find, following conventional lore, that the sum rule
relating the interquark potential to the field energy in the continuum
formulation would be violated in the lattice regularized theory. The
implications of
this result are discussed.

Consider the ground state energy $\hat E_0$ of a quark-antiquark
pair separated by a distance $\hat R$. Quantities denoted with a
``hat'' will always be understood to be measured in
units of the lattice spacing. The energy $\hat E_0$ can be
calculated from the expectation value of the Wilson loop with
spatial and temporal extension $\hat R$ and $\hat T$,
respectively:
$$\hat E_0=-\lim_{\hat T\to\infty}{1\over\hat T}\ln<W(\hat R,\hat T)>
\ \ .\eqno(1)$$
On an isotropic lattice $<W(\hat R,\hat T)>$ is calculated with the action
$$S=\hat\beta({\cal P}_\tau+{\P}_s)\eqno(2)$$
where, for $SU(N)$, $\hat\beta$ is defined in terms of the bare coupling
constant by $\hat\beta={2N\over g^2_0}$,
and ${\P}_\tau,{\P}_s$ denote the contributions of the
time-like and space-like plaquette variables:
$$\eqalign{
{\P}_\tau&=\sum_n\sum_{\mu\not=4}\left[1-{1\over{2N}}Tr(U_{\mu 4}
(n)+U^\dagger_{\mu 4}(n))\right],\cr
{\P}_s&=\sum_n\sum_{i>j}\left[1-{1\over{2N}}Tr(U_{ij}
(n)+U^\dagger_{ij}(n))\right].\cr}\eqno(3)$$
Here $i,j$ label the spatial directions, and $U_{\mu\nu}(n)$ is the
lattice variable associated with a plaquette located in the
$\mu\nu$-plane at the lattice site $n$.

The lattice energy $\hat E_0$, defined by (1), is a function of $\hat R$
and $\hat\beta$. Since the self-energy contributions to
$\hat E_0$, associated with the quark and antiquark, do not depend
on $\hat R$, they can be eliminated by considering the difference
$\hat E_0(\hat R,\hat\beta)-\hat E_0(\hat R_0,\hat\beta)$,
where $\hat R_0$, is some reference $q\bar q$-separation.  Then the subtracted
$q\bar q$
potential is given by
$$[V(\hat R,\hat\beta)]_{subtr}=-\lim_{\hat T\to\infty}{1\over \hat T}[\ln
<W(\hat
R,\hat T>]_{subtr} \ ,\eqno(4a)$$
where
$$<W(\hat R,\hat T)>={\int DU W(\hat R,\hat T)e^{-\hat\beta({\P}_\tau
+{\P}_s)}\over \int DUe^{-\hat\beta({\P}_\tau+{\P}_s)}}.\eqno(4b)$$
{}From here on we will always assume that such a subtraction has been
carried out, and shall drop the subscript "subtr" for simplicity.
Following [4] we now take the derivative of (4a) with
respect to $\hat\beta$. One then obtains
$${\partial\hat V(\hat R,\hat\beta)\over \partial \hat\beta}=\lim_{\hat
T\to\infty}{1\over\hat T}<{\P}_\tau + {\P}_s>_{q\bar q-0},\
\eqno(5a)$$
where the expectation value $<O>_{q\bar q-0}$ is defined generically by
$$<O>_{q\bar q-0}={<W(\hat R,\hat T)O>\over<W(\hat R,\hat T)>}
-<O>.\eqno(5b)$$
Hence the $\hat\beta$-derivative of the potential is expressed in
terms of correlators of plaquette variables with the Wilson loop.
These are the correlators which have been measured in Monte
Carlo simulations to determine the spatial distribution of the energy
density in a flux tube connecting a quark and antiquark [6].

In the limit $\hat T\to\infty$, the rhs of (5a) can be
further simplified. Since for $\hat T\to\infty$ the
expectation value of a plaquette variable is invariant under
a ``time'' translation, we have that
$$<{\P}_\sigma>_{q\bar q-0}\Buildrel \approx \over {\hat T\to\infty}\hat T
<{\P}_\sigma'>_{q\bar q-0},
\eqno(6)$$
where ${\P}_\sigma'$ is given by an expression of the form (3), with $n$
running over the lattice sites on the fixed time slice. With a Wilson
loop extending from $n_4=-{\hat T\over2}$ to $n_4={\hat T\over2}$,
this time slice is conveniently chosen to be the
$n_4=0$ plane. Then
$$\hat\beta{\partial\hat V(\hat R,\hat\beta)\over\partial\hat\beta}
\approx \hat\beta <{\P}_\tau' + {\P}_s'>_{q\bar q-0}
\to  a\sum_{\vec x}a^3{1\over2}
<{\vec E^2}(\vec x)+{\vec B}^2(\vec x)>_{q\bar q-0},\eqno(7)$$
where in the last step we have taken the naive continuum
limit. Here $-{1\over2}{\vec E}^2$ and ${1\over 2}{\vec B}^2$ are the
Minkowski energy densities of the chromoelectric and chromomagnetic
fields expressed in terms of the euclidean fields.

We next use the renormalization group to cast the lhs of (7)
in a form involving the potential and its derivative with respect
to $\hat R$. In the limit of vanishing lattice spacing
"$a$" we have that
$${1\over a}\hat V\left({R\over a},\hat\beta(a)\right)
\Buildrel \longrightarrow \over {a\to0} V(R),\eqno(8)$$
where $V(R)$ is the physical interquark potential, and the
behaviour of $\hat\beta(a)$ as a function of the lattice
spacing is given, close to the continuum limit, through
the renormalization group relation
$$a={1\over\Lambda_L}\left({2Nb_0\over\hat\beta}\right)^{-b_1/(2b^2_0)}
e^{-{\hat\beta\over 4Nb_0}}.\eqno(9)$$
Here $b_0$ and $b_1$ are given by
$$b_0={11N\over48\pi^2};\quad b_1={34\over3}\left({N\over
16\pi^2}\right)^2.\eqno(10)$$
The invariance of the lhs of (8) with regard to changes in the lattice
spacing leads to
$${\partial\hat\beta\over\partial ln a}{\partial\hat V(\hat R,\hat
\beta)\over \partial \hat\beta}=\hat V(\hat R,\hat\beta)
+\hat R{\partial\hat V(\hat R,\hat\beta)\over\partial\hat R} \ , \eqno(11)$$
where it is understood that this expression is to be evaluated for
$\hat R={R\over a}$, and with $\hat\beta(a)$ determined by (9). Making use of
the relation (11),
equation (7) takes the following form close to the continuum limit
$$\hat V(\hat R,\hat\beta)+\hat R{\partial\hat V(\hat R,\hat\beta)
\over\partial\hat R}={\partial\hat\beta\over\partial ln a}
<{\P}_\tau' + {\P}_s'>_{q\bar q-0}\eqno(12)$$
In the case of a confining potential, $\hat V(\hat R,\hat\beta)=
\hat\sigma(\hat\beta)\hat R$, this equation reduces to
$$2\hat\sigma(\hat\beta)\hat R={\partial\hat\beta\over
\partial ln a}<{\P}_\tau' + {\P}_s'>_{q\bar q-0}\eqno(13)$$
which coincides with that obtained in [3] by making use of the relations
(3.7), and the equation following it in that reference. The second term
appearing on the lhs of (12), which gives rise to the factor of
two in (13), has been missed in [4]. Hence the Michael action
sum rule is incorrect, as was also recently observed in [5], where the authors
derive this sum rule within the continuum formulation of QCD.

A second sum rule, relating the interquark
potential to the Minkowski field energy, can be obtained by requiring
that a lattice regularization involving different lattice spacings
in the temporal and spatial directions should lead to the same
physical potential as that computed from an isotropic lattice. A
similar argument has been used in [3] for the string tension
computed either from a time-like or a space-like Wilson loop,
leading to a coupling constant sum rule which will play a key role in
our discussion. On an anisotropic lattice the action involves two
couplings, $\hat\beta_\tau$ and $\hat \beta_s$ [1-3], associated
with the time-like and space-like plaquette contributions:
$$S = \hat\beta_\tau{\P}_\tau+\hat\beta_s{\P}_s. \eqno(14)$$
These couplings are conventionally considered to be functions of the
spatial lattice spacing $a$ and the anisotropy parameter $\xi=a/a_\tau$,
where $a_\tau$ is the lattice spacing in the temporal direction.
They are usually parametrized as follows [2,3]
$$\hat\beta_s ={2N\over g^2_s(a,\xi)}\xi^{-1}, \ \ \
\hat\beta_\tau={2N\over g^2_\tau(a,\xi)}\xi,\eqno(15)$$
where
$g^2_s(a,1)=g^2_\tau(a,1)=g^2_0(a)$,
and where the explicit dependence on $\xi$ is chosen in such a way
that the $\xi$ dependence of $g_\sigma(a,\xi), \sigma=s,\tau$,
arises only from quantum corrections. In the weak coupling limit
$\hat\beta_\tau(a,\xi)$ and $\hat\beta_s(a,\xi)$ can
be related to the bare coupling $\hat\beta(a)$ on an isotropic
lattice [2,3] by
$$\eqalign{
{1\over\xi}\hat\beta_\tau(a,\xi)=&\hat\beta(a)
+2N c_\tau(\xi)+{\cal O}({\hat\beta}^{-1}),\cr
\xi\hat\beta_s(a,\xi)=&\hat\beta(a)
+2N c_s(\xi)+{\cal O}({\hat\beta}^{-1}),\cr}\eqno(16)$$
where the $\xi$-dependence of the functions $c_\sigma(\xi), \sigma=\tau,s$,
have
been studied in detail in [3].

With the action (14), the lattice potential computed from the expectation
value of the Wilson loop becomes a function
of $\hat R, \hat\beta_s$, and $\hat\beta_\tau$. We now require
that in the continuum limit $a\to 0, a_\tau\to0, \xi=a/a_\tau$ fixed,
the physical potential, $V\sim{1\over a_\tau}\hat V$,
should not depend on the choice of $\xi$, if the couplings
$\hat\beta_s$ and $\hat\beta_\tau$ are tuned with $a$
appropriately. Hence for $a\to 0$ we have that
$${d\over d\xi}\left[{\xi\over a}\hat V\left({R\over a},
\hat\beta_\tau(a,\xi),\hat\beta_s(a,\xi)\right)\right]=0 \ . \eqno(17)$$
Noting that
$${\partial \hat V\over\partial \beta_\sigma}= <{\P}_\sigma'>
_{q\bar q-0};\quad \sigma=\tau,s,\eqno(18)$$
one finds, upon carrying out the differentiation (17),
and then returning to the isotropic lattice $\xi=1$, that
$$\hat V(\hat R,\hat\beta)=-\left[\left({\partial\hat\beta_\tau
\over\partial\xi}\right)<{\P}_\tau'>_{q\bar q-0}+
\left({\partial\hat\beta_s
\over\partial\xi}\right)<{\P}_s'>_{q\bar q-0}\right]_{\xi =1}\eqno(19)$$
Here $\hat V(\hat R,\hat\beta)$ is the potential in lattice
units computed on an isotropic lattice. Again it is understood that
this relation only holds for $\hat R=R/a,\beta_\sigma=\beta_\sigma
(a,\xi)$, in the continuum limit. Defining
$$\eta_{\pm}={1\over 2}\left[\left({\partial\hat\beta_\tau\over\partial\xi}
\right)
_{\xi=1} \pm \left({\partial\hat\beta_s\over\partial\xi}\right)
_{\xi=1}\right],\eqno(20)$$
expression (19) can be written in the form
$$\hat V(\hat R,\hat\beta)=\eta_- <-{\P}_\tau' +
{\P}_s'>_{q\bar q-0}
-\eta_+<{\P}_\tau' +
{\P}_s'>_{q\bar q-0} \ ,\eqno(21)$$
where the expectation values are computed on an isotropic lattice.
Making use of the action sum rule (12) one obtains
$$\hat V(\hat R,\hat\beta)+\eta_+{\partial ln a\over\partial\hat\beta}
\left[\hat V(\hat R,\hat\beta)+\hat R{\partial\hat V
(\hat R,\hat\beta)\over\partial\hat R}\right]
=\eta_-
<-{\P}_\tau' + {\P}_s'>_{q\bar q-0}\eqno(22)$$
We emphasize that so far our discussion has not involved any perturbative
arguments. We now want to interprete the rhs of (22) in terms of a continuum
physical
observable. It is at this (and only this) stage where, following standard
lore, we make use of
the weak coupling approximations (16).
{}From the definition (20) and the relations (16) one finds that for $a\to0$
$$\eta_-\approx \hat\beta(a)+N[c_\tau'(1)-c_s'(1)]
\Buildrel \longrightarrow \over {\hat\beta\to\infty} \hat\beta(a) \
,\eqno(23)$$
where $c'_\sigma$ is the derivative of $c_\sigma$.
Hence close to the continuum limit
$$\hat V(\hat R,\hat\beta)+\eta_+{\partial ln a\over\partial\hat\beta}
\left[\hat V(\hat R,\hat\beta)+\hat R{\partial\hat V\over\partial\hat R}(\hat
R,
\hat\beta)\right]\approx\hat\beta<-{\P}_\tau'+{\P}_s'>_{q\bar q-0},
\eqno(24)$$
where, as explained before, ${\P}_\sigma'$ $(\sigma=\tau,s)$
denotes the contributions to the action of
plaquettes with base on a fixed time slice.
In the continuum limit $\hat\beta(-{\P}_\tau'+{\P}_s')$
can be identified with the Minkowski field energy measured in lattice units.
Thus for $a\approx 0$,
$$\hat\beta<-{\P}_\tau'+{\P}_s'>_{q\bar q-0}\approx a\sum_{\vec x}
a^3{1\over2}<-{\vec E}^2(\vec x)+{\vec B}^2(\vec x)>_{q\bar q-0} \ ,
\eqno(25)$$
where $(-{1\over2}{\vec E}^2)$ and ${1\over 2}{\vec B}^2$ are the
electric and magnetic contributions to the (Minkowski) field
energy densities expressed in terms of the euclidean fields. Now
the field energy, after subtracting the self energy contributions, is
expected to be
related to the potential by
$$V(R)={1\over2}\int d^3x
<-{\vec E}^2(\vec x)+{\vec B}^2(\vec x)>_{q\bar q-0} \ .
\eqno(26)$$
Hence on the lattice ${1\over a}\hat V\left({R\over a},\hat\beta(a)\right)$
should also match the total field energy for $a \to 0$. Thus if (24)
and (25) holds, then we
are led to conclude that
the second term appearing on the lhs of (24) should vanish in this
limit. This is
the case if either $\hat V+\hat R{\partial\hat V\over \partial \hat R}=0$,
$\partial ln a/\partial\hat\beta=0$, or $\eta_+=0$. In the
first case $\hat V(\hat R,\hat\beta)$ has the form $\hat V\sim
\alpha(\hat\beta)/\hat R$, and hence does not allow for a confining
potential. The second possibility is clearly excluded for $SU(N)$
gauge theories. Hence we conclude that in a confining theory like
QCD, $\eta_+$ would have to vanish for $\hat\beta \to \infty$. But
according to (20) and (15),
$$\eta_+=N\left\lbrace\left({\partial g^{-2}_\tau\over\partial\xi}
\right)_{\xi=1}+\left({\partial g^{-2}_s\over\partial\xi}
\right)_{\xi=1}\right\rbrace$$
Hence the above line of reasoning would lead
to the conclusion that the quantity appearing in curly brackets
should vanish in the limit $\hat\beta \to \infty$. This disagrees with
the result obtained in
ref. [3], where this
quantity was shown to be given by $b_0$, defined in (10). The author
was led to this result by requiring the invariance of the string
tensions computed from space-time and space-like Wilson loops
under changes in the anisotropy parameter $\xi$. On an isotropic lattice the
two string tensions extracted in this way should coincide. By making
further use of the action sum rule, Karsch was led to the above
conclusion.

Using the value $\eta_+=Nb_0$ obtained in [3], the approximation for
$\eta_-$ given
in (23), and the one-loop approximation $\partial ln a/\partial\hat\beta
=-{1\over 4Nb_0}$, one finds that, for a confining potential, (24) reduces
to
$$\hat\sigma\hat R\approx2\hat\beta<-{\P}'_\tau+{\P}'_s>_{q\bar q-0}\ ,
\eqno(27)$$
which, with the identification (25), would violate the energy sum rule (26)
by a
factor of two. We emphasize that the origin of the factor two in (27)
is the second term appearing on the rhs of
(21). With $\eta_-\approx \hat\beta=2N/g^2_0$, eq. (21) takes
the form
$$\hat V(\hat R,\hat\beta)\approx{2N\over g^2_0}\Bigl\lbrace
<-{\P}'_\tau+{\P}'_s>_{q\bar q-0}
-{g^2_0\over 2N}\eta_+<{\P}'_\tau +{\P}'_s>_{q\bar q-0}\Bigr\rbrace \ .
\eqno(28)$$
The contribution proportional to $<{\P}'_\tau+{\P}'_s>_{q\bar q-0}$
is absent in the expression obtained in [4].
The energy sum rule would have the naively expected form,
$\hat V = \hat\beta<-{\P}'_\tau+{\P}'_s>_{q\bar q-0}$, if one
neglects this {\it apparent} perturbative contribution.
The action sum rule (13) for a confining potential,
together with the value $\eta_+ = Nb_0$ taken from [3],
however tells us that this contribution is about ${1\over2}\hat V$
for a confining potential.

Summarizing, our above discussion has shown, that if the weak coupling
relations (16) are used to identify the rhs of (22) with the field
energy in the continuum limit, then the implementation of the energy sum rule
(26) on the
lattice demands that $\eta_+$ vanishes in this limit, which contradicts
the coupling constant sum rule obtained in $\lbrack 3 \rbrack$.
This latter sum rule has however been derived without invoking
perturbation theory.  We are therefore rather
tempted to conclude that the weak coupling argument leading to (24) with the
identification (25) is inadequate, and that the lattice expression for the
field energy is that defined by the right hand side of (21). This stands
in sharp
contrast to the standard belief, that the field energy is related to
$<-{\P}'_\tau+{\P}'_s>_{q\bar q-0}$, which as we have seen only holds for a
$1/R$-potential. For a linearly rising potential, on the other hand, the
contribution
proportional to $<{\P}'_\tau+{\P}'_s>_{q\bar q-0}$ is essential, and implies
that
the rhs of (22) approaches in the continuum limit only one half
of the field energy stored in the flux tube connecting the
quark-antiquark pair, irrespective of the $SU(N)$ gauge group. Clearly
the lattice energy sum rule deserves further
investigations. In particular it would be of interest to determine the
$\hat\beta$-dependence
of $\eta_-$ from Monte Carlo calculations of Wilson loops on anisotropic
lattices.
\bigskip\bigskip

\centerline{\bf Acknowledgments}
\smallskip\noindent
We are very grateful to R. Banerjee, H.G. Dosch, O. Nachtmann, and in
particular
to I.O. Stamatescu for useful discussions and valuable comments.
\bigskip
\centerline{\bf REFERENCES}
\bigskip\noindent
[1] J. Kuti, J. Polonyi and  K. Szlanchanyi, Phys. Lett. {\bf 98B} (1981) 199;
J. Engels, F. Karsch, I. Montvay and H. Satz, Phys. Lett. {\bf 101B} (1981) 89
\smallskip\noindent
[2] A. Hasenfratz and P. Hasenfratz, Nucl. Phys. {\bf B193} (1981) 210
\smallskip\noindent
[3] F. Karsch, Nucl. Phys. {\bf B205} [FS 5](1982) 285
\smallskip\noindent
[4] C. Michael, Nucl. Phys. {\bf B280} [FS 18] (1987) 13
\smallskip\noindent
[5] H.G. Dosch, O. Nachtmann and M. Rueter, Heidelberg preprint HD-THEP-95-12
\smallskip\noindent
[6] For some early work see: R. Sommer, Nucl. Phys. {\bf B291} (1987) 673;
ibid {\bf B306} (1988) 180.
R.W. Haymaker and J. Wosiek, Phys. Rev. Rapid Comm. {\bf D36} (1987) 3297;
R.W. Haymaker, Y. Peng, V. Singh and
J. Wosiek, Nucl. Phys. (Proc. Suppl.) {\bf B17} (1990) 558;
R.W. Haymaker, V. Singh and J. Wosiek,
Nucl. Phys. (Proc. Suppl.) {\bf B20} (1991) 207; R. Haymaker and
J. Wosiek, Phys. Rev. {\bf D 43} (1991) 2676
\smallskip\noindent

\end